# Étude comparée de quatre logiciels de gestion de références bibliographiques libres ou gratuits

Gérald Kembellec
gerald.kembellec@cnam.fr
Laboratoire Paragraphe
Université Paris 8
2 rue de la Liberté
93526 Saint-Denis

Claire Scopsi
claire.scopsi@cnam.fr
Laboratoire Dicen
Conservatoire National des Arts et Métiers
2 rue Conté
75003 Paris

**Résumé**

Cet article résulte d'une réflexion sur lesoutils de gestion de références bibliographiques, particulièrement ceux proposés sous une forme libre ou gratuite.Depuis 2007,l'interaction des outils de rédactionavec leséditeursbibliographiques évolue rapidementmaispar le passé les logiciels libres ont pu souffrir de la comparaison en termes d'ergonomie ou d'usageavec l'offre propriétaire. Cepanoramafonctionnel et technique approfondi des solutions libres ou gratuites actuellesrésulte de la comparaisondes logicielsJabRef, Mendeley Desktop, BibDesk et Zoteromenée en janvier 2012 par deux enseignants chercheurs au sein de l'institut national français des techniques de la documentation (INTD).

**Introduction**

L'étape de la collecte d'information est cruciale dans le processus d'écriture scientifique ou technique et elle peut être grandement facilitée par le choix d'un outil numérique dédié performant. Un logiciel de gestion de références bibliographiques (LGRB) est un programme destiné à établir, trier et publier des listes de citations relatives à des revues, des articles, des sites web, des ouvrages, principalement dans le cadre de publications scientifiques. Ces logicielssont notamment utilisés par les étudiants, enseignants et chercheurs de l'enseignement supérieur ainsi que par les bibliothécaires et documentalistes des milieux académiques. Ils sont généralement composés d'une interface de gestion reliée à une base de données qui peut être alimentée de différentes façons et se distinguent souvent par leur capacité à importer et exporter les différents formats informatiques reconnus.

Bien sûr une collection de notices bibliographiques peut être conçue à l'aide d'un éditeur de texte classique. Quel avantage trouve-t-onalors à utiliser un outil dédié ? Tout d'abord ce type de logiciel garantit l'intégrité et la cohésion des références dans la base de données car il fonctionne avec des règles strictes fondées sur les normes en vigueur et évite d'oublier des ponctuations ou des informations obligatoires. Par ailleurs, avec le temps, une collection s'accroît et la maintenir manuellement à l'aide d'un éditeur de texte devient complexe (il n'est pas rarequ'une thèse, par exemple, comporteune bibliographie de plusieurs milliers de lignes). Enfin les outils bibliographiques intègrent des fonctionnalités de classement, de recherche, voire d'annotation dans la collection dont l'intérêt n'est pas négligeable.

Dans ce panorama nous confronterons les diverses fonctionnalités de ces outils aux besoins des usagers ce quisituenotre étude dans la continuité des travaux de Zweifel(Zweifel 2008) et de ceux de Masur(Masur 2009). Il est intéressant en effet de rendre compte de l'avancement de leurs fonctionnalités depuis ces dernières années et d'ajouter, comme le suggérait Carole Zweifel, des « retours d'expérience d'utilisations réelles ». Le respect des standards et la compatibilité sontà analyser avec un soin particulier car, au-delà des fonctionnalités d'usage, la qualité que perçoit l'utilisateur de ces logiciels dépend des formats d'import et d'export. De son côté, l'étude de Masur concluait que les logiciels de gestion de références bibliographiques libres sont encore méconnus du grand public et peinent à se faire une place au sein du marché.

**Une offre diversifiée**

Les logiciels de bibliographie les plus connus et usités sont RefWorks et Endnote(Dell'Orso 2010). Ces produits de qualité, issus de sociétés éditrices traditionnelles, sont malheureusement d'un coût élevé et leur licence d'exploitation limite strictement leur usage et leur diffusion. De plus, leurs codes sources (manière dont les logiciels sont écrits) sont souvent protégés. Il est donc impossible à l'utilisateur, pour des raisons contractuelles (la licence d'utilisation) et matérielles (le blocage de l'accès aux sources), de programmer des extensions[i]. Cela interdit donc l'adaptation de l'outil aux besoins spécifiques d'un usager ou d'une équipe de recherche.

Pour ces raisons, nous examinons deux types de LGRB : d'une part l'offre libre, c'est-à-dire une gamme de logiciels dont le code, ouvert, est modifiable et dont la licence autorise un usage et une diffusion gratuits, d'autre part des logiciels dont la licence reste propriétaire mais dont l'usage est gratuit. Ces deux familles de logiciels sont en effet particulièrement prisées dans les milieux universitaires.

Dans le premier temps de notre étude, nous présentons les conditions de réalisation des essais et discutons les critères que nous considérons les plus pertinents pour l'évaluation de ces produits. Ces critères se regroupent en quatre domaines :
- A. les critères « non techniques » concernant les licences et les conditions de développement du logiciel,
- B. les plateformes techniquessupportées,
- C. la compatibilité et le respect des formats standards,
- D. les fonctionnalités d'usage.

Dans un deuxième temps nous livrons les résultats des essais de quatre logiciels : JabRef, Mendeley Desktop, BibDesk et Zotero. En effet, une étude récente montre que ces produits sont prisés par les étudiants, enseignants-chercheurs et documentalistes de l'enseignement supérieur(Kembellec 2012).

La troisième partie évalue l'adéquation de ces logiciels aux attentes de trois catégories d'utilisateurs : les étudiants de 3e cycle, les chercheurs et les bibliothécaires ou documentalistes. L'étude précédemment citée montre en effet qu'EndNote (dans le monde propriétaire), Zotero et Mendeley(dans le monde libre) sont particulièrement adaptés aux documentalistes tandis queles pratiquesdes étudiants et enseignants-chercheurs sont influencées par les usages d'écriture scientifique. Sans entrer dans les détails, en sciences, techniques et médecine l'usage conjoint de Zotero avec JabRef ou Mendeley est fréquent tandis qu'en sciences humaines et sociales Zotero est très largement utilisé, parfois en collaboration avec Mendeley.

## Les conditions de l'évaluation

Les critères d'évaluation

Les critères fonctionnels sont au cœur de cette évaluation mais ils ne doivent pas occulter les critères liés à la licence qui déterminent les usages autorisés, les conditions d'évolution du produit et ceux liés à l'architecture qui conditionnent leur exploitation technique et leur pérennité. En effet de par leur nature les logiciels de bibliographie concernent des usages au domicile ou en milieu professionnel (laboratoire, université, bibliothèque) et peuvent être manipulés par des usagers plus ou moins aguerris. Enfin la sécurité des données ne doit pas être perdue de vue : certaines références bibliographiques accompagnent un chercheur pendant de longues années.

### *Mode de développement et droits d'usages*

*Licence libre ou non ?*

Les logiciels libres sont le plus souvent gratuits, mais les logiciels gratuits ne sont pas forcément libres.Les logiciels sous licence libres respectent les règles fondamentales émises par Richard Stallman (libre accès au code source pour l'étudier et le modifier, libre usage y compris en situation professionnelle ou commerciale, libre diffusion - y compris payante - du logiciel lui-même ou d'une version modifiée) et sont particulièrement adaptés au domaine de l'enseignement et de la recherche. Prônant le respect des standards, ils sont appréciés dans les laboratoires en raison leur adaptabilité ; ils peuvent en outre être diffusés auprès de collaborateurs et d'étudiants ou sur les postes d'un Service Commun de Documentation en toute légalité et sans paiement de redevance.

Il existe également, dans le domaine bibliographique des logiciels dont les licences ne sont pas libres mais dont l'usage est gratuit. Nous nousintéressons également à cesproduits tout en émettant les réserves d'usage car,si leur utilisation individuelle est libre de droits, leur diffusion (par exemple sur les postes d'un centre de ressource) est le plus souvent interdite, et l'accès et la modification du code source sont rarement autorisés. Il est donc important de prendre connaissance des termes de la licence et de ne pas déduire de sa gratuitéqu'un logiciel est d'utilisation libre.

*Caractéristiques de la structure de développement*

Les logiciels libres peuvent être publiés par une société éditrice ou par une communauté libre. Comme pour les logiciels propriétaires il est préférable de vérifier, en consultant les sites et forums de la communauté, la vitalité de la structure de développement. En effet, une communauté active a une plus grande capacité à produire de nouvelles versions et garantit un logiciel pérenne.

Si l'image d'un développement collaboratif est souvent associée au libre et fortement défendue par Richard Stallman(Stallman 1992), la réalité est plus nuancée (Jullien et Zimmerman 2005; Bergquist et Ljungberg 2001). Il n'est pas rare qu'un éditeur ou une communauté limitent ou refusent les contributions des usagers pour préserver l'unité et la qualité du produit. Dans ce cas il sera toujours possible à l'usager de modifier le logiciel et de diffuser ces modifications de son côté sans que ces évolutions soient réintégrées au logiciel d'origine.

Certains logiciels libres dont le développement est très contrôlé sont structurés de façon à accepter des modules additionnels (greffons ou add-on) développés par les usagers sans être intégrés dans le code du logiciel principal. Ces modules ne sont pas toujours validés par la communauté ou l'éditeur et parfois diffusés sous d'autres plateformes. Il faut donc les choisir et les tester avec soin mais ce procédé offre des possibilités d'extension du logiciel intéressantes.

*Architecture logicielle*

*Lourd / Léger / Hybride*

Le terme« clientlourd »fait référenceà une interface homme-machine individuelle permettant de piloter une base de connaissances locale ou distante. Le code de l'application s'exécute sur une machine locale ou sur un serveur d'applications après avoir été chargé en mémoire. Ce code peut nécessiter une compilation préalable. Nous nous intéresserons dans un premier temps à cesLGRBque l'on installe sur une station de travail. Dans un deuxièmetemps nous abordons les « *Rich Internet Applications* (RIA) »qui sont hébergées sur un serveur web et utilisées au travers d'un navigateur (on parle aussi de « client léger »).

Enfin il existe un mode « hybride » carcertains LGRBoccupent une place à part à mi-chemin entre le client lourd et le RIA. Il s'agit le plus souvent de greffons pour navigateurs web qui prennent en chargele glanage d'informations par détection de métadonnées embarquées dans une page web ou le moissonnageet l'import massif depuis un dépôt de documents.

Cette question a une incidence sur le déploiement et l'utilisation des outils. Le logiciel lourd nécessitant une installation poste par poste, le déploiement et la mise à jour sur le lieu de travail doivent être organisés, surtout dans les structures qui n'autorisent pas les droits d'installation aux collaborateurs. Les RIA n'ont pas cet inconvénient puisqu'il suffit de connaître l'URL de l'application et de disposer d'un identifiant pour l'utiliser depuis n'importe quel poste connecté à internet. C'est un avantage pour les utilisateurs nomades qui utilisent différents ordinateurs (les étudiants dans une bibliothèque ou une salle de ressources informatiques par exemple).

Les logiciels hybrides comportent un élément actif installé sur le poste de travail. Comme les logiciels lourds,cet élément doit être déployé localement et le niveau de sécurité du navigateur doit être paramétré de façon à autoriser son exécution.

*Stockage des données*

Le mode de stockage des données importées peut égalementdifférer. Certains logiciels stockent les informations dans un fichier dit « à plat », c'est-à-dire dans un fichier texte structuré selon un format spécifique. Les différents éléments sont alors séparés par des caractères déterminés (point-virgule, tabulation…) ou encadrés par des « balises » XML. D'autres outils utilisent une base de données relationnelle (MySQL par exemple), ce qui permet de manipuler les données avec un logiciel d'administration de bases de données standard. Cela peut faciliter l'exploitation des données (la sauvegarde), leur versement dans un logiciel documentaire ou leur migration vers un autre outil bibliographique.

*Interopérabilité et respect des standards*

La compatibilité des LGRBavec les plateformes de diffusion, c'est-à-dire leur capacitéà importer et exporter des bibliographies normées est un critère essentiel de l'évaluation. Ainsi, bien qu'invisibles pour l'utilisateur, les structures des données sont repérables par des logiciels adaptés(Poupeau, 2011)mais aussi par les moteurs de recherche[ii]. Ces métadonnées sont dites « exposées » ou « glanables » s'il existe une ou plusieurs méthodes de valoriser les contenus pour un système informatique tiers. Les formats d'exposition web de notices testés sont les suivants :

- Le BibT$_E$X, format de bibliographie pour l'écriture scientifique, associé à l'écriture au format LaT$_E$X ;
- Le Dublin Core sous forme de métadonnées aux formats RDFa ouMicroformats(Poupeau 2011) ;
- Le format COinS(Trainor et Price 2010; Van de Sompel 2001) de la norme Z39-88[iii].
- Les informations DOI(Lupovici).

*Les formats*

Le format d'import de notices : le logiciel doit être capable d'importer des bibliographies ou des collections complètes déjà pré formatées par d'autres LGRB sans perte ni corruption de données.
Le format d'export de notices : il est de même indispensable de pouvoir exporter tout ou partie de la base de notices dans un ou plusieurs formatsen respectant les normes en vigueur.
La compatibilité avec les traitements de texte : pour compléter les fonctionnalités d'un LGRB, il est important de pouvoir verser directement des références bibliographiques dans son traitement de texte.

Les LGBR proposent parfois la gestion des doublons (qui garantit la qualité de la base)et la détection d'erreur de typage documentaire (repérage d'éléments manifestement manquants pour un type de document, comme le numéro de page dans un article de revue).

*Fonctionnalités d'usage*

Pour retrouver les références correspondant aux citations et auteurs l'utilisateur peut souhaiter annoter sa base de citations ou effectuer des classements par sujet. Une base indexée offre la possibilité d'effectuer des filtres de recherche.

Pour des travaux collaboratifs, la possibilité de partager sa bibliothèque de citations est une option indispensable. Cette fonctionnalité est parfois offerte sous la forme d'un répliquât de la base de connaissance sur un serveur internet. Cette option permet également de sauvegarder sa base bibliographique.

**Les logiciels évalués**

JabRef

### *Présentation de JabRef*

JabRef est une interface graphique de gestion qui permet de maintenir exclusivement des bibliographies au format BibT$_E$X. Ce logiciel est multiplateforme car développé en java. Cette programmation autorise donc une compatibilité maximale en terme de systèmes d'exploitation. Ainsi le portage de ce logiciel se fera aussi bien sur MAC OS, Windows que Linux.

Le projet JabRef est né de la rencontre de Morten O. Alver et NizarBatada. Le premier avait développé JBibtexManager et le second BibKeeper. La première version de JabRef, née de la fusion des deux logiciels, a été diffusée en 2003. Le nom JabRef signifie *J* pour Java, *a* pour Alver, *b* pour Batada et *Ref* pour Référence. Ce projet libre a rapidement connu un franc succès dans la sphère de la recherche universitaire. Les éditeurs de systèmes d'exploitation libres l'ont intégré comme un *package* disponible pour de nombreuses versions de Linux depuis 2007.

Mendeley Desktop

### *Présentation de Mendeley Desktop*

Mendeley Desktop est, comme le substantif anglais « *Desktop* » l'indique, un logiciel installé localement sur le disque dur de l'ordinateur. Cette application est, comme les autres, spécialisée en gestion bibliographique et permet également de partager des références à travers son interface web.

Mendeley Ltd. est une société londonienne. La première mouture du logiciel éponyme a été publiée mi-2008. Cette solution est gratuite, mais pas « *Open Source* » etl'extension de stockage en ligne est payante. De nouvelles fonctionnalités sont prévues sous forme d'extensions payantes, comme MendeleyInstitutional Edition pour les établissements d'enseignement supérieur.

BibDesk

### *Présentation de BibDesk*

BibDesk est une solution logicielle libre et gratuite qui permet de créer, gérer et utiliser des données bibliographiques au format BibT$_E$X. Ce programme doté d'une interface graphique claire et efficace n'est disponible que sur les ordinateurs Apple.

Lancé publiquement en 2002, BibDesk est en développement continu depuis cette date. Bibdesk est un exemple de logiciel libre supporté par de nombreux contributeurs grâce à la plateforme Sourceforge. A l'origine le développeur en était Michael McCracken et une grande partie du code a ensuite été écrite par Adam Maxwell et Christiaan Hofman. Disponible directement à partir de SourceForge, il est actuellement livré avec la distribution de T$_E$XMacT$_E$X.

Zotero, le module du navigateur Firefox

*Présentation de Zotero*

Zotero est un module logiciel (« plug-in »), développé par le *Center for History and New Media* de la *George MasonUniversity*. Zotero est à part dans le panorama des LGRB car ce module s'exécute au sein du navigateur Firefox dont il est une composante optionnelle. Zotero permet de rentrer manuellement des notices au sein d'une base locale, de les classer et de les exporter sous différents formats bibliographiques.

La première version deZoteroa étémise à la dispositiondes utilisateurs en octobre 2006en tant que greffon logiciel pourle navigateurweb Firefox(Laskowski). Depuis, une très importante communauté internationale s'est regroupée pour participer à la création de styles bibliographique. Depuis janvier 2011 il existe une version de type logiciel lourd de Zotero, « *standaloneZotero* », comportant les mêmes fonctionnalités sans nécessiter l'usage de Firefox.

Protocole de test

Lorsque cela était possible ces logiciels ont été testés dans les environnements suivants :

- Mac OS X sur un MacBook Pro équipé de MS Office 2008 et 2011LyX, Texstudio, TexWorks et OpenOffice.

- Un ordinateur portable récent équipé de Windows 7, MS Office 2007, LyX et OpenOffice.

Pour les logiciels compatibles avec le glanage de notices bibliographiques sur les pages web, nous avons déposé un fichier HTML bien formé avec des notices au format « web de données » sur un site Internet de test. Grâce à cette URL, nous pourrons juger de la compatibilité du logiciel avec la navigation en mode glanage d'informations. Nous avons également prévu de tester la compatibilité des logiciels avec les métadonnées associées aux documents. Pour y parvenir nous proposerons aux logiciels de charger un panel de documents scientifiques au format PDF. Le corpus est composé de fichiers PDF bien formés et correctement annotés, grâce à des métadonnées au format XMP[iv].

**Résultats**

Résultats de JabRef

| Licence et Modèle économique | Type de logiciel | Détection de doublons et d'erreurs | Annotation, classement, recherche | Détection de notices intégrées web / PDF | Sauvegarde en ligne / partage | Communication avec un éditeur de texte |
|---|---|---|---|---|---|---|
| Libre, Gratuit, | Logiciel lourd | Oui / Oui | Oui / Oui / Oui | Oui / Oui | Oui / Non | LyX, Kile, Word (après manipulation) |

*Tableau 1. Résumé de l'évaluation de JabRef*

*Formats d'import de JabRef*

JabRef importe principalement les notices grâce à des fichiers BibT$_E$X et BibT$_E$XML mais aussi RIS et Refer. Bien que ce logiciel s'appuie sur une interface graphique, son moteur peut être alimenté en ligne de commande. Cette opération peut trouver son intérêt lors de traitements par lot en *shell script*. Les adeptes de la ligne de commande, tels les amateurs de Linux, seront particulièrement intéressés par cette possibilité de charger des bibliographies issues de divers fichiers aux formats différents.

*Formats d'export de JabRef*

Les principaux formats d'export normés supportés sont le RIS, le MODS, le RDF (cf. (Poupeau; Poupeau 2011)). Outre ces quelques formats d'export classiques (cf. Figure 1), JabRef offre la possibilité de se connecter à une base de données et d'exporter sa bibliographie sous MySQL. Cette possibilité est particulièrement attractive après un moissonnage ou un glanage intensif d'information avec export en format normé. Le processus est donc presque automatisé entre la navigation avec le module de Firefox Zotero et la génération d'une base de connaissances de tous les articles intéressants sur un sujet. Il suffit pour cela de configurer une connexion à un serveur MySQL (qu'il soit local ou distant).

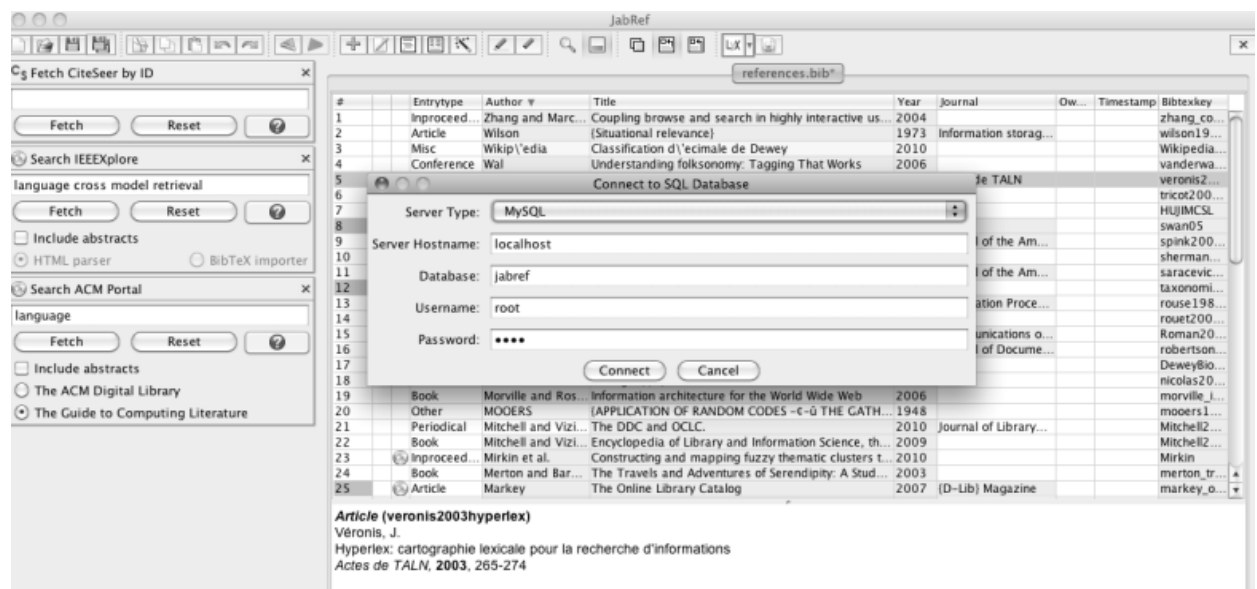

*Figure 1. Export de Jabref vers une Base MySQL*

*Fonctionnalités de JabRef*

Pour les rédacteurs en sciences dites « dures », l'utilisation de LaT$_E$X et l'usage de BibT$_E$X avec un fichier bien formalisé forment un pré requis implicite pour soumettre des articles dans la plupart des revues. JabRef offre l'opportunité de formaliser des bibliographies de manière très stricte. En effet, si les champs obligatoires ne sont pas renseignés ou sont mal renseignés, le logiciel le signale de manière claire par un point rouge. Un autre avantage de ce logiciel est qu'il est gratuit et s'installe facilement. L'article de Toulin, Chazelas, et Palencia (2008) présente égalementJabRef comme compatible avec plusieurs traitements de texte, dont MS Word via le format RTF[13]. La suite de bureautique la plus compatible est OpenOffice et sa version redevenue libreLibreOffice. JabRef est capable de générer une bibliographie au format OpenDocument (.ods), le format du tableur de LibreOffice,pour ensuite l'importerdans

le traitement de texte. Ce n'est que depuis janvier 2011 et la version 2.7 bêta du logiciel qu'un début de compatibilité avec Word de MS Office se concrétise autrement que par un export RTF[v]. Il semble qu'avec quelques manipulations de configuration JabRef soit compatible avec MS Word [8] directement en mode insertion.

Une autre fonctionnalité décrite par les auteurs du logiciel est que JabRef permet d'inscrire ses propres métadonnées dans les fichiers PDF grâce à la technologie XMP. Grâce à ces métadonnées les chercheurs peuvent échanger des fichiers annotés et enrichis de mots clés et de commentaires. Ainsi, au sein du logiciel, les chercheurs ont accès non seulement au document primaire et à sa notice, mais aussi aux annotations et métadonnées ajoutées par leurs collègues.

Le point faible de ce logiciel réside dans sa médiocre interaction avec les outils de traitements de texte classiques. En effet une phase d'export en XML OpenDocument est nécessaire pour importer les citations dans MS Word, orcette manipulation n'est pas anodine pour un utilisateur non aguerri. Cependant, comme son usage est clairement orienté pour le couple BibT$_E$X, LaT$_E$X, l'inconfort reste mineur. Son point fort est la souplesse d'utilisation, l'adaptabilité, mais également la rigueur de création de fichier BibT$_E$X.

Résultats de Mendeley Desktop

| Licence et Modèle économique | Type de logiciel | Détection de doublons et d'erreurs | Annotation, classement, recherche | Détection de notices intégrées web / PDF | Sauvegarde en ligne / partage | Communication avec un éditeur de texte |
|---|---|---|---|---|---|---|
| Propriétaire, Gratuit, avec extensions payantes | Logiciel lourd, RIA, Hybride | Oui / Non | Oui / Oui / Oui | Oui / Oui | Oui / Oui | MS Word(après manipulation) |

*Tableau 2. Résumé de l'évaluation de Mendeley*

***Formats d'import de Mendeley Desktop***

Mendeley Desktop permet d'importer directement les métadonnées depuis les documents, notamment les fichiers PDF. Les autres formats d'import sont le Ovid, le RIS, le BibT$_E$X, le EndNote XML, le TXT (texte brut), mais aussi, et c'est un avantage, la base Sqlite du moduleZotero de Firefox.

***Formats d'export de Mendeley Desktop***

Mendeley propose de créer des bibliographies sous divers formats tant libres que propriétaires. Les plus usités sont le BibT$_E$X, RIS, texte brut, XML, Zotero (zotero.sqlite), PDF. Il est possible de copier/coller des citations vers BibT$_E$X directement de Mendeley par le presse-papier du système d'exploitation. Mendeley propose également un module de rédaction de citation à intégrer directement dans Word et Open Office sous forme d'une barre d'outils.

***Fonctionnalités de Mendeley Desktop***

Ce logiciel offre la possibilité de partager des références et les fichiers associés au sein de librairies partagées entre plusieurs utilisateurs d'un même groupe. Dans ce cadre l'espace de stockage est limité à 500 méga-octets par compte. Les groupes ne peuvent pas accueillir plus de 10 participants par collection partagée ou *« Shared Collection »*. Mais dans ce cadre la détection des doublons n'est pas gérée. Des options payantes permettent d'augmenter l'espace de stockage. Mendeley permet de partager ses références et les documents attachés ainsi que les métadonnées personnalisées. Ce logiciel offre une gestion des fichiers de type PDF performante, capable de lire les métadonnées des documents. On peut donc générer les notices bibliographiques depuis les documents (voir **Erreur ! Source du renvoi introuvable.**). Cette commodité nécessite tout de même une relecture systématique. En effet selon les éditeurs de textes et les modèles fournis par les organisateurs de conférences l'organisation des métadonnées peut être altérée. Mendeley est compatible avec les logiciels MS Word et OpenOffice ; nous avons pu tester le module pour Word 2008 et 2011 qui autorise l'insertion de références dans le texte et l'édition de la bibliographie.

L'extension« *bookmarklet »,* une fois intégrée au navigateur, permet de détecter les notices au format COinS et de les importer dansMendeley, à la condition d'avoir un compte en ligne.

D'un point de vue fonctionnel, un élément très appréciable est la possibilité d'annoter un document PDF au sein même du logiciel qui réalise la bibliographie. Dans l'exemple explicité par la capture (cf. **Erreur ! Source du renvoi introuvable.**2), le document de l'auteur Dinet a pu être annoté (1) et réutilisé ultérieurement. En effet au cours de la lecture nous avions pu surligner (2) les passages clés et annoter le document grâce à Mendeley Desktop. Par la suite la note apparait comme un *post-it* flottant dans le document, près de son point d'ancrage (3).

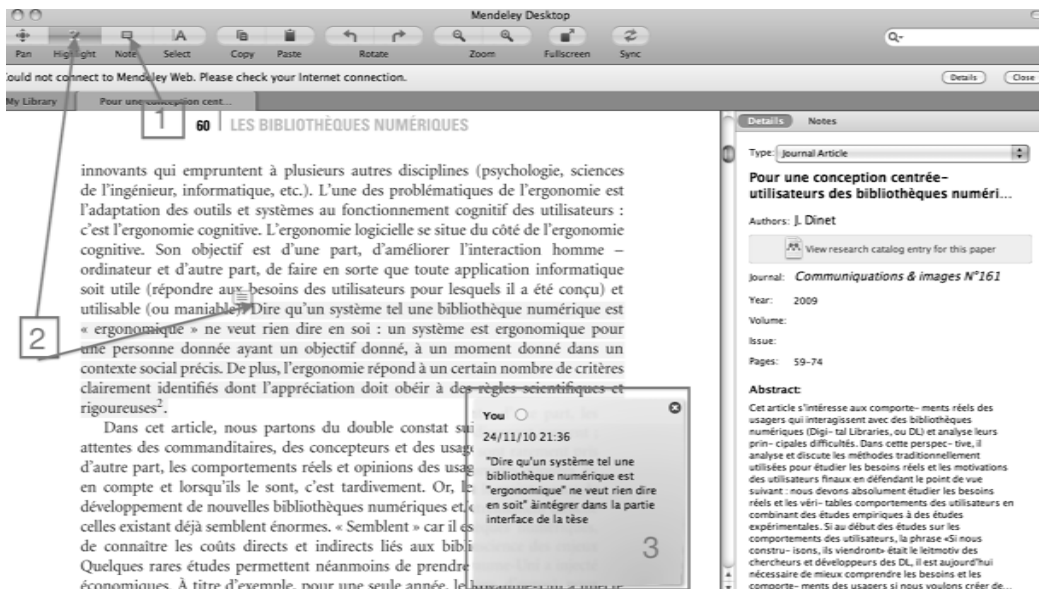

*Figure 2. Annotation de document grâce à Mendeley.*

Résultats de Bibdesk

| Licence et Modèle économique | Type de logiciel | Détection de doublons et d'erreurs | Annotation, classement, recherche | Détection de notices intégrées web / PDF | Sauvegarde en ligne / partage | Communication avec un éditeur de texte |
|---|---|---|---|---|---|---|
| Libre, Gratuit. | Logiciel lourd | Oui / Non | Oui / Oui / Oui | Oui / Oui | Non, Non | LyX, Kile, Word |

*Tableau 3. Résumé de l'évaluation de BibDesk*

*Plateforme d'exécution et format d'import unique*

L'éditeur graphique BibDesk n'est disponible que sur plateforme Apple. Il est dédié aux bibliographies LaT$_E$X c'est-à-dire au format BibT$_E$Xet ne gère donc que les fichiers à extension *.bib*. L'interface est simple, la Figure 3 montre l'édition d'un article avec sur la partie de droite les hyperliens vers doi (2) et la bibliothèque numérique (1) où l'article est officiellement référencé. Les auteurs sont également clairement identifiés en bas à droite (3). La partie de gauche est réservée aux métadonnées de l'article dont les mots clés qui revêtent une importance particulière (4).

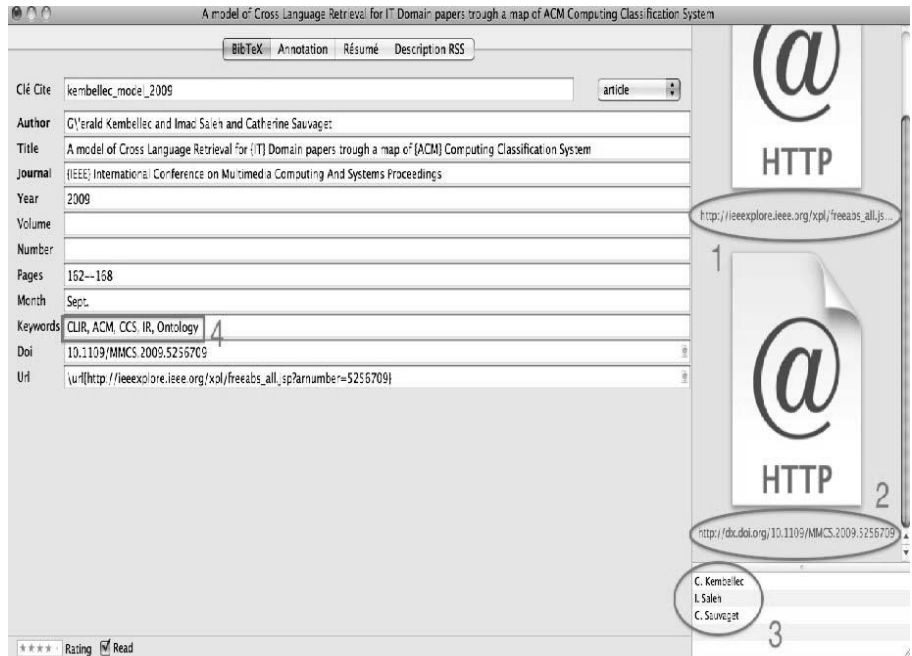

*Figure 3. Interface d'édition de référence bibliographique de BibDesk.*

*Formats d'export de BibDesk*

Le premier et principal format d'export de BibDesk est bien sûr le BibT$_E$X. Utiliser un logiciel limite les possibilités d'erreur à la compilation du document LaT$_E$Xqui, par exemple,exige que tous les caractères accentués soient reformatés pour être valides etcette option de « nettoyage » n'est pas négligeable lors de la compilation d'une thèse comportant plus d'une centaine de citations. Un temps conséquent de débogage peut ainsi être gagné. BibDesk propose également d'exporter des notices bibliographiques ou des bibliographies complètes dans les formats RIS, et COinS dans un site Web par BibDesk.

*Fonctionnalités complémentaires de BibDesk*

Ce logiciel ne s'arrête pas à l'édition des notices contenues dans une bibliographie au format LaT$_E$X ; il intègre en outre un système d'auto complétion des mots clés et des fonctionnalités de classement «intelligent» grâce à ces mêmes mots clés. Les politiques de classement s'effectuent également à partir de filtres sur des opérateurs de type « est » ou « contient » et peuvent s'appliquer sur n'importe quel champs BibT$_E$X, par exemple l'auteur, le titre, ou les mots clés. Dans l'exemple de la Figure 4, les éléments 1 représentent les champs « tester » avec l' « opérateur » 2 et les « arguments » 3. Cela peut être traduit en langage naturel par « création d'un groupe filtrant les documents comportant les mots clés

ACM et CCS. »

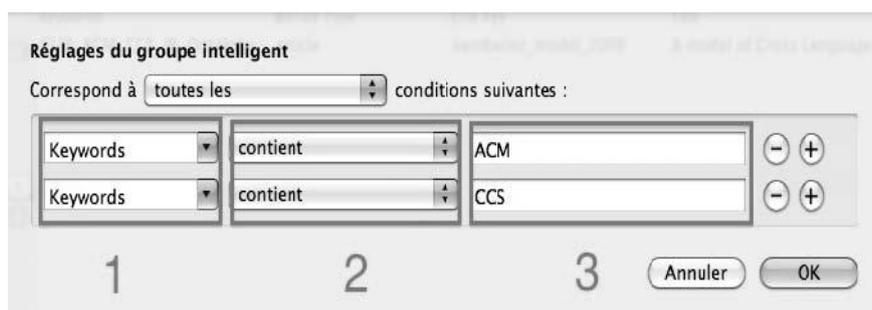

*Figure 4. Fonctionnalité de recherchedans la base bibliographique de BibDesk*

BibDesk possède son propre moteur d'indexation auquel il est possible d'appliquer des règles de classement en « groupes intelligents ». Le moteur de recherche de BibDesk permet d'analyser la structure des fichiers bibliographiques au format BibT$_E$X ainsi que bon nombre de bases de connaissances prédéfinies. Parmi ces bases de connaissances, citons :

- des sites dont l'accès en consultation est gratuit comme ACM (méta données et résumés en accès libre), arXiv, CiteULike, Google Scholar, IACR (Cryptologie);
- des sites accessibles par abonnement comme IEEE Xplore, MathSciNet, Project Euclid, SpringerLink, Zentralblatt Math;
- mais aussi les sites respectant les normes bibliographiques : BibT$_E$X, COinS, HCite. Cette option permet d'utiliser BibDesk comme un navigateur web dédié à la recherche scientifique et d'insérer à la volée les notices nécessaires à l'établissement d'une bibliographie.

BibDesk possède une interface de navigation web qui propose la détection de notices bibliographiques au format OpenURL. Il est ainsi possible d'insérer directement les citations dans la base de notices d'un simple clic sur le bouton « importer ».

Depuis janvier 2011 la communauté Bibdesk a écrit un module qui permet d'utiliser BibDesk avec le traitement de texte MS Word pour insérer dynamiquement des citations et générer la bibliographie d'un document.

*Formats d'import de Zotero*

Bien sûr la capacité de Zotero à détecter un document ou une notice dépendra de l'exposition des métadonnées. Il faut à minima que le document soit décrit de manière correcte en Dublin Core sur la page XHTML. Les résultats visibles à l'œil, c'est-à-dire les liens vers les ressources, sont ainsi accompagnés par des métadonnées descriptives normalisées. Zotero est à même de comprendre ces métadonnées et de les proposer à l'usager, soit en import massif, soit en ne sélectionnant que les documents intéressant le chercheur. Les notices glanées peuvent ensuite être éditées, triées et classées. Un moteur de recherche intégré permet d'accéder intuitivement au catalogue ainsi réalisé.

Résultats de Zotero

| Licence et Modèle économique | Type de logiciel | Détection de doublons et d'erreurs | Annotation, classement, recherche | Détection de notices intégrées web / PDF | Sauvegarde en ligne / partage | Communication avec un éditeur de texte |
| --- | --- | --- | --- | --- | --- | --- |

| Libre, Gratuit avec extensions payantes | Logiciel hybride, RIA, Logiciel lourd | Oui / Non | Oui / Oui / Oui | Oui / Oui | Oui / Oui | MS Word , OO |

*Tableau 4. Résumé de l'évaluation de Zotero*

### Formats d'export de Zotero

Zotero permet l'export de tout ou partie de sa base de citations aux formats RIS, BibT$_E$X, RDF, Refer et MODS. Il est possible de copier une référence dans le presse-papier pour ensuite la coller dans son traitement de texte favori en tant qu'élément de bibliographie ou texte formaté. Il existe un module d'intégration pour MS Word qui permet l'import formaté depuis Zotero.

### Fonctionnalités de Zotero

Notons que Zotero permet la génération de fichiers au format BibT$_E$X. Il semble que le gros avantage de Zotero soit de pouvoir détecter intelligemment aussi bien les références bibliographiques que les documents eux-mêmes. Zotero permet donc une chaine de production quasi automatisée allant de la détection ou l'import jusqu'à la génération de bibliographies en de nombreux styles et sous de nombreux formats. Il est compatible avec Word qui est l'autre outil de production scientifique et il est même possible d'exporter une citation bibliographique par un simple « cliquer déplacer » vers un éditeur ou un client de messagerie. Sous Word 2008 et 2011 (MAC OS), un portage a été proposé pour insérer des citations depuis Zotero et générer dynamiquement la bibliographie avec quelques styles.

Ce module possède également un moteur de recherche intégré permettant de parcourir les références enregistrées dans la base de l'utilisateur. Le réel intérêt de Zotero est de proposer la possibilité de détecter sur le web des ressources documentaires de différents types et d'enregistrer en masse des notices bibliographiques grâce aux métadonnées exposées par le site web.

Ici, à l'issue d'une requête sur Google Scholar, la réponse est constituée d'éléments bibliographiques détectables séparément. L'icône Zotero de la barre de navigation signale la présence de multiples entrées. En cliquant sur cette icône, une fenêtre d'alertepropose d'enregistrer une ou plusieurs notices dans la base (cf. Figure 5).

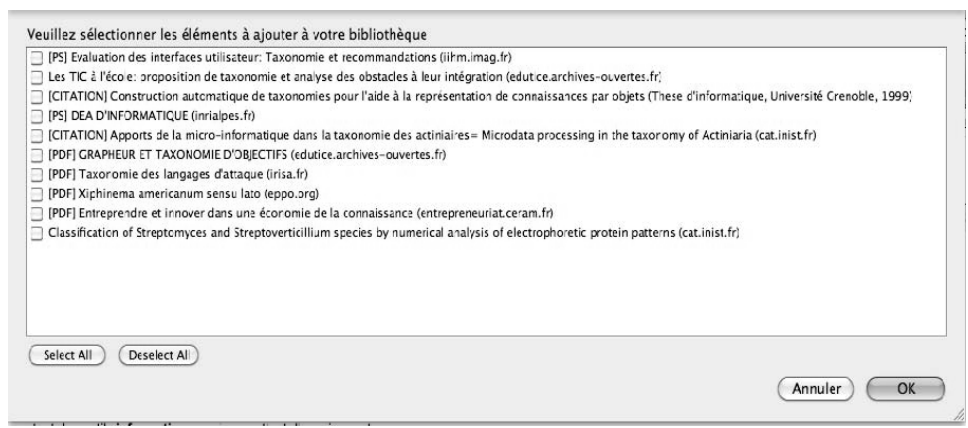

*Figure 5.Alerte de Zotero proposant un import sélectif de citations.*

Zotero permet également au navigateur de détecter et d'enregistrer un élément isolé, comme la page personnelle d'un chercheur offrant la lecture de son dernier article. S'il s'agit d'un document unique décrit sur une page Web le module précisera le type du document (un article, un livre ou une vidéo) lors de sa détection, en affichant une icône à droite dans la barre de navigation (par exemple un livre, un chapitre, un article d'encyclopédie sera symbolisé par un petit livre bleu).

L'interface de Zotero est intégrée à Mozilla Firefox et peut s'activer à tout moment en cliquant sur le logo Zotero en bas à droite du navigateur. Dans la Figure , sur la gauche en 1, les collections sont classées dans la « Bibliothèque » qui représente l'ensemble des notices bibliographiques. Au centre, en 2, il est possible de visualiser l'ensemble des articles composant une collection. En 3, un encart permet de visualiser et d'éditer les informations formant la notice relative à un document. L'onglet 4 offre la possibilité d'afficher les mots clés potentiellement associés à la fiche. Comme ces champs sont rarement renseignés il est possible de les choisir soi-même. Cette option est précieuse pour ceux qui désirent filtrer les articles par mots clé grâce au moteur intégré en 5. En cas d'export d'une bibliographie les mots clés seront également exportés.

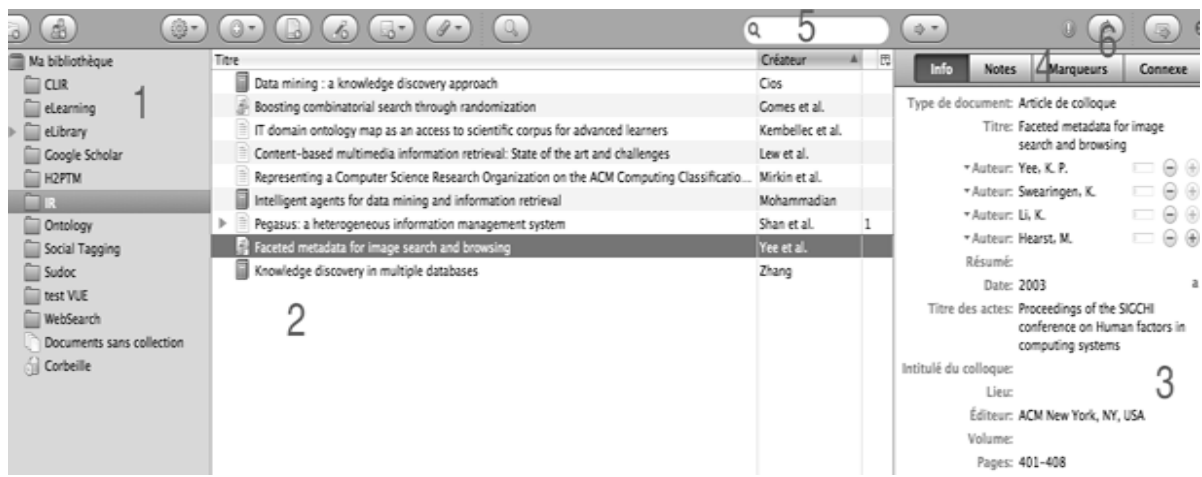

*Figure 6. Interface de Zotero*

La force de ce logiciel est d'autoriser la création et l'export de bibliographies aux formats d'éditeurs de textes les plus répandus. De plus Zotero offre une mise en forme bibliographique adaptable au domaine de recherche de l'utilisateur, voire même au type de revue, grâce à un vaste de panel de styles bibliographiques pour les traitements de texte. Depuis janvier 2012Zotero existe également sous forme de logiciel indépendant ce qui n'oblige plus à utiliser Firefox.

**Quel LGRB pour quel usage ?**

Pour connaître les pratiques en matière de création assistée de bibliographies nous nous référons à une enquête réalisée en décembre 2011(Kembellec 2012) auprès d'un panel d'utilisateursde 195 personnes. Le groupe est constitué de conservateurs et de documentalistes de Services Communs de Documentation d'universités parisiennes, d'enseignants chercheurs en mathématiques, informatique, sciences de la communication et même d'un maître de conférences en lettres ayant fait sa thèse en LaT$_E$X (bien que ces pratiques soient moins répandues hors des sciences « dures » et des domaines techniques). Ils ont été interrogés sur la manière dont ils écrivaient leurs bibliographies et les réponses montrent que les usages et pratiques divergent selon les profils d'utilisateurs mais aussi selon le type de sciences de rattachement.

En STM, BibTeX est souvent imposé comme format de bibliographie par les revues et éditeurs scientifiques. Pour gérer leur bibliographie les chercheurs en sciences dures utilisent pour plus de la moitié d'entre eux un logiciel dédié, souvent JabRef ou Mendeley, en collaboration avec Zotero. Un usage conjoint de Zotero avec un autre outil permet de repérer et d'enregistrer les notices depuis internet puis de gérer la bibliographie avec un logiciel dédié. Cependant, en médecine, le format bibliographique Ovid associé à MS Word pour l'écriture, géré par Mendeley, est une alternative reconnue. Nous proposons donc en sciences dures (sauf médecine) l'usage de JabRef ou BibDesk si l'on préfère écrire sous LaTEX. Dans le cas d'une pratique d'écriture sous Ms Word, il faut préférer Zotero ou Mendeley.

En SHS, les documents scientifiques étant essentiellement rédigés avec des traitements de texte classiques, les outils dédiés à LaT$_E$X sont hors de propos. Même s'ils commencent à devenir compatibles, mieux vaut privilégier des logiciels éprouvés tels que Zotero ou Mendeley. Cependant, ceux qui présentent des travaux à la mise en page avancée avec des polices de caractère intégrant des caractères de langues mortes ou rares préféreront parfois LaT$_E$X. Dans ce cas précis, assez rare, l'usage de JabRef ou BibDesk peut être utile.

Dans les centres de documentation, les usagers de l'information bibliographique sont souvent déjà équipés d'un LGRB, souvent payant, comme EndNote. Cela ne les empêche pas de connaître et d'utiliser Zotero en complément.

L'étude montre que toutes les catégories d'usagers (enseignants, étudiants, documentalistes) utilisent souvent Zotero en association avec un autre logiciel. Parmi les logiciels libres ou gratuitsZotero a semble-t-il le monopole de détection des notices bibliographiques et Mendeley celui de la gestion des bibliographies. Cependant avec l'arrivée de Zorero « *standalone* », nous allons peut-être assister à l'émergence d'un monopole.

**Conclusion**

De plus en plus de solutions libres et gratuites de gestion de bibliographies sont désormais opérationnelles etles logiciels propriétaires n'ont plus le monopole de la qualité en ce domaine. Il existe donc un vrai choix de logiciels selon le système d'exploitation et le format bibliographique. Comme le faisait remarquer Carole Zweifel, ces logiciels présentent autant, sinon plus, d'innovations fonctionnelles que les produits payants (Zweifel 2008). Car, avec l'évolution des usages tels le web 2.0 et le web sémantique, les créateurs de ces produits en sont les premiers usagers. Parmi toutes ces applications libres les chercheurs devraient trouver un logiciel offrant une compatibilité satisfaisante en matière d'import, d'export, de détection et de pratique collaborative.

Il est tout à fait possible d'utiliser les compatibilités de formats entre les logiciels pour effectuer veille technologique, recherche et formatage normé depuis divers logiciels. Par exemple, le greffon de Firefox Zotero est particulièrement indiqué pour glaner des notices exposées sur internet. Ensuite, les fiches seront exportées au format BibT$_E$X ou RIS dans JabRef, Bibdesk ou Mendeley. Le fichier d'export en .bib ou autre contenant la bibliographie pourra également être exporté vers une interface internet riche (RIA) qui sera accessible depuis n'importe quel poste nomade équipé d'internet et d'un navigateur. Notons que presque tous les logiciels lourds (Mendeley) ou hybride (Zotero) possèdent une plateforme de sauvegarde ou de synchronisation en ligne qui peut parfois également servir de base communautaire. Le chaînage logiciel requiert une certaine expertise rapidement acquise par la manipulation. Mais pour ceux qui souhaitent un minimum de manipulation et une tranquillité d'esprit sur la qualité de leur bibliographie, JabRef est le bon compromis car la possibilité de détecter et manipuler les métadonnées des articles au format PDF rend son usage simple et convivial. De plus les formats d'import-export étant devenus compatibles avec tous les environnements, son usage ne se cantonne plus à créer des bibliographies pour les éditeurs OpenOffice et LaT$_E$X.

À la question : peut on trouver une alternative sérieuse aux logiciels de gestion bibliographique « professionnels » payants ?, la réponse est claire : oui, bien sûr. Il existe maintenantdes programmes de qualité adaptés à chaque spécialité scientifique, mais aussi en accord avec les sensibilités individuelles face aux différents modèles économiques de distribution.

**Notes de fin**

[i] Une extension est un « greffon » logiciel, c'est-à-dire un morceau de programme venant élargir les fonctionnalités du logiciel.

[ii] La nouvelle mouture de l'algorithme de Google « Panda »se base principalement sur ces données descriptives pour juger de la qualité d'une page web.

[iii] Voir le blog d'Emmanuelle Bermès de la BnF :
http://www.figoblog.org/document1131.php, accédé le 1er septembre 2012.

[iv] Cet acronyme signifie *eXtensibleMetadata Platform*, plate-forme de métadonnées extensible. Il s'agit d'un standard proposé par Adobepour intégrer des métadonnées à des fichiers de type PDF.

[iv] http://jabref.sourceforge.net/revisionhistory.php, accédé le 1$^{er}$ septembre 2012.